# Hadronic Rare $B$ Decays via Exchange or Annihilation Diagrams

**Zhi-zhong Xing** [1]
*Sektion Physik, Theoretische Physik, Universität München,*
*Theresienstrasse 37, D-80333 München, Germany*

**Abstract**

The two-body mesonic $B$ decays induced only via a single $W$-exchange or annihilation quark diagram, such as $B^-_u \rightarrow D^{(*)-}_s K^{(*)0}$ and $\bar{B}^0_d \rightarrow D^{(*)+}_s K^{(*)-}$, are analyzed in the factorization approximation. We estimate the branching ratios for those transitions into two pseudoscalar mesons, and find them to be negligibly small. The significant effect of final-state rescattering is illustrated by taking $B^-_u \rightarrow D^- \bar{K}^0$ for example.

[1] Electronic address: Xing@hep.physik.uni-muenchen.de

As a result of the large data sample of weak $B$ transitions collected on the $\Upsilon(4S)$ resonance by CLEO and ARGUS collaborations [1], the hadronic decays of $B$ mesons appear to be a valuable window for determining the quark mixing parameters, probing the origin of $CP$ violation and investigating the nonperturbative confinement forces. Further experimental efforts towards the above physical goals, such as the program of CLEO III and the construction of KEK and SLAC $B$-meson factories, are underway.

The dynamics of exclusive hadronic $B$ decays, in particular those via the $W$-exchange or annihilation quark diagrams, is not yet well understood. The decay rates of $W$-exchange and annihilation transitions are usually argued to be negligibly small due to the suppression of helicity and (or) formfactors [2], however, a solid justification of this argument is necessary in both theory and experiments. Current data have given upper bounds on some of the $W$-exchange or annihilation decay modes of $B$ mesons, e.g., $\mathrm{Br}(B^-_u \rightarrow D^{*-}_s K^0) < 1.2 \times 10^{-3}$ and $\mathrm{Br}(\bar{B}^0_d \rightarrow D^{*+}_s K^{*-}) < 1.2 \times 10^{-3}$ [1]. A better understanding of such processes is possible in the near future, with the accumulation of larger data samples.

As a preliminary step towards comprehensive studies of the hadronic rare $B$ transitions via $W$-exchange and annihilation diagrams, this short note concentrates on the two-body mesonic decays. We first survey all possible decay modes of this nature by use of a complete quark-diagram scheme, and then estimate branching ratios for those channels into two pseudoscalar mesons. Finally the significant effect of final-state rescattering is illustrated by taking $B^-_u \rightarrow D^- \bar{K}^0$ for example.

According to the topology of lowest-order electroweak interactions with QCD effects included, all two-body mesonic $B$ decays can be graphically described in terms of ten distinct quark diagrams [3] [2]. In the assumption of no final-state rescattering or channel mixing, it is possible to survey those "pure" decay modes induced only by a single quark graph. We find that the following neutral $B$ decays occur solely through the $W$-exchange diagram (see Fig. 1(a) for illustration):

$$\begin{array}{lll} \bar{B}^0_d & \longrightarrow & D^{(*)+}_s K^{(*)-} \; , \;\; D^{(*)-}_s K^{(*)+} \; ; \\ \bar{B}^0_s & \longrightarrow & D^{(*)-} M^+ \; , \;\; D^{(*)0} M^0 \; , \;\; D^{(*)+} M^- \; , \;\; \bar{D}^{(*)0} M^0 \; , \end{array} \tag{1}$$

[2]Note that the popular six-graph scheme [4] does not include the quark diagram for the color-matched electroweak penguin transitions and those for the decays where one (or both) final-state meson(s) must be the flavor singlet(s). For a detailed discussion, see Ref. [3].

where $M$ denotes an $I=1$ light unflavored meson like $\pi$, $\rho$ or $a_1$ [1]. For two-body charged $B$ transitions, the following decay modes occur only via the annihilation diagram (see Fig. 1(b) for illustration):

$$\begin{aligned} B^-_u & \longrightarrow D^{(*)-}\bar{K}^{(*)0} \,, \quad D^{(*)-}_s K^{(*)0} \,; \\ B^-_c & \longrightarrow K^{(*)-}K^{(*)0} \,, \quad K^{(*)-}M^0 \,, \quad \bar{K}^{(*)0}M^- \,, \quad M^-M^0 \,. \end{aligned} \tag{2}$$

In experiments, some of the above processes have been searched for, and upper limits to the branching ratios of $\bar{B}^0_d \rightarrow D^{(*)+}_s K^{(*)-}$ and $B^-_u \rightarrow D^{(*)-}_s K^{(*)0}$ have been obtained [1, 5]. Nevertheless, there has not been any preliminary estimation of the decay rate for any of these modes.

The effective weak Hamiltonian responsible for the decays in Eqs. (1) and (2) is given by

$$\mathcal{H}_{\rm eff} \;=\; \frac{G_F}{\sqrt{2}} V_{cb}V^*_{uq}\left[c_1\;(\bar{q}u)_{V-A}(\bar{c}b)_{V-A} \;+\; c_2\;(\bar{c}u)_{V-A}(\bar{q}b)_{V-A}\right] \;+\; (u \Leftrightarrow c) \;+\; {\rm H.c.}\,, \tag{3}$$

where $q=d$ or $s$, $V$ represents the Cabibbo-Kobayashi-Maskawa (CKM) matrix, $c_1$ and $c_2$ are two Wilson coefficients at the scale $O(m_b)$. Assuming a generic decay mode $B_\delta(b\bar{\delta}) \rightarrow X(\alpha\bar{\gamma}) + Y(\gamma\bar{\beta})$ as illustrated in Fig. 1, one can factorize its amplitude $\langle XY|\mathcal{H}_{\rm eff}|B_\delta\rangle$ into a product of three terms: the CKM factor, the combination of Wilson coefficients and the matrix element of color-singlet currents. For example,

$$\langle D^-\bar{K}^0|\mathcal{H}_{\rm eff}|B^-_u\rangle \;=\; \frac{G_F}{\sqrt{2}}\;a_1\;(V_{ub}V^*_{cs})\;\Omega^{D^-\bar{K}^0}_{scu} \tag{4a}$$

or

$$\langle D^-_s K^+|\mathcal{H}_{\rm eff}|\bar{B}^0_d\rangle \;=\; \frac{G_F}{\sqrt{2}}\;a_2\;(V_{ub}V^*_{cd})\;\Omega^{D^-_sK^+}_{ucd}\;, \tag{4b}$$

where $a_1 \equiv c_1 + c_2/3$, $a_2 \equiv c_2 + c_1/3$, and the hadronic matrix elements are obtained from the definition

$$\Omega^{XY}_{\alpha\beta\delta} \;\equiv\; \langle XY|(\bar{\alpha}\beta)_{V-A}|0\rangle\langle 0|(\bar{\delta}b)_{V-A}|B_\delta\rangle\;. \tag{5}$$

Subsequently we treat $a_1$ and $a_2$ as free parameters, in order to phenomenologically accommodate the contribution of color-octet currents which has been neglected in the above naive factorization approximation [6, 7]. The matrix elements $\Omega^{XY}_{\alpha\beta\delta}$ can be Lorentz-invariantly decomposed in terms of the decay constants and formfactors, however, many difficulties exist in evaluating the relevant annihilation formfactors.

For simplicity and illustration, here we only calculate $\Omega^{XY}_{\alpha\beta\delta}$ for the case that both $X$ and $Y$ are pseudoscalar mesons. Following the work of Bernabéu and Jarlskog [8], we obtain

$$\Omega^{XY}_{\alpha\beta\delta} \;=\; {\rm i}\;\frac{m_X - m_Y}{m_X + m_Y}\left[(m_X+m_Y)^2 - m^2_{B_\delta}\right] f_{B_\delta}F^{\rm a}_+(m^2_{B_\delta})\;, \tag{6}$$

where $f_{B_\delta}$ is the decay constant of $B_\delta$ meson, and $F_+^{\rm a}(m_{B_\delta}^2)$ is the annihilation formfactor. The perturbative QCD calculation gives $F_+^{\rm a}(m_{B_\delta}^2) = {\rm i}16\pi\alpha_s f_{B_\delta}^2/m_{B_\delta}^2$ [9], which is primarily absorptive. In estimating the branching ratios of $B_\delta \rightarrow XY$, we take $\alpha_s(m_b) = 0.20$, $a_1 = 1.15$ and $a_2 = 0.26$ [5]. The central values of meson masses can be found from Ref. [1]. The average lifetimes of $B_u$, $B_d$, $B_s$ and $B_c$ mesons are taken to be $1.54\times 10^{-12}$s, $1.50\times 10^{-12}$s, $1.34\times 10^{-12}$s and $0.5\times 10^{-12}$s respectively [1, 10]. Considering the constraints of unitarity on the CKM matrix $V$ [11], we adopt $|V_{ud}| = 0.9744$, $|V_{cs}| = 0.9734$, $|V_{us}| = |V_{cd}| = 0.22$, $|V_{cb}| = 0.04$ and $|V_{ub}| = 0.08|V_{cb}|$. We also input $f_{B_u} = f_{B_d} = 0.196$ GeV, $f_{B_s} = 0.212$ GeV, $f_{B_c} = 0.48$ GeV and $M_{B_c} = 6.25$ GeV [12, 10]. The numerical results are listed in Table 1.

| Decay mode | Quark diagram | CKM factor | Wilson factor | Branching ratio |
|---|---|---|---|---|
| $B_u^- \rightarrow D^- \bar{K}^0$ | annihilation | $V_{ub}V_{cs}^*$ | $a_1$ | $8.1\times 10^{-9}$ |
| $B_u^- \rightarrow D_s^- K^0$ | annihilation | $V_{ub}V_{cd}^*$ | $a_1$ | $4.2\times 10^{-10}$ |
| $\bar{B}_d^0 \rightarrow D_s^+ K^-$ | $W$-exchange | $V_{cb}V_{ud}^*$ | $a_2$ | $6.5\times 10^{-8}$ |
| $\bar{B}_d^0 \rightarrow D_s^- K^+$ | $W$-exchange | $V_{ub}V_{cd}^*$ | $a_2$ | $2.1\times 10^{-11}$ |
| $\bar{B}_s^0 \rightarrow D^+ \pi^-$ | $W$-exchange | $V_{cb}V_{us}^*$ | $a_2$ | $1.2\times 10^{-8}$ |
| $\bar{B}_s^0 \rightarrow D^0 \pi^0$ | $W$-exchange | $V_{cb}V_{us}^*$ | $a_2$ | $1.2\times 10^{-8}$ |
| $\bar{B}_s^0 \rightarrow D^- \pi^+$ | $W$-exchange | $V_{ub}V_{cs}^*$ | $a_2$ | $1.5\times 10^{-9}$ |
| $\bar{B}_s^0 \rightarrow \bar{D}^0 \pi^0$ | $W$-exchange | $V_{ub}V_{cs}^*$ | $a_2$ | $1.5\times 10^{-9}$ |
| $B_c^- \rightarrow K^- K^0$ | annihilation | $V_{cb}V_{ud}^*$ | $a_1$ | $6.3\times 10^{-9}$ |
| $B_c^- \rightarrow \pi^- \pi^0$ | annihilation | $V_{cb}V_{ud}^*$ | $a_1$ | $1.1\times 10^{-7}$ |
| $B_c^- \rightarrow K^- \pi^0$ | annihilation | $V_{cb}V_{us}^*$ | $a_1$ | $6.7\times 10^{-6}$ |
| $B_c^- \rightarrow \bar{K}^0 \pi^-$ | annihilation | $V_{cb}V_{us}^*$ | $a_1$ | $6.4\times 10^{-6}$ |

Table 1: Typical examples of weak $B$ decays into two pseudoscalar mesons via a single $W$-exchange or annihilation diagram.

From Table 1 we observe that all 12 decay modes have negligibly small branching ratios in the context of the factorization approximation and the formfactor model used above. The suppression of decay rates mainly arises from the smallness of $F_+^{\rm a}(m_{B_\delta}^2)$ and $|V_{ub}|$. For those $W$-exchange induced channels, the smaller Wilson factor $a_2$ also suppresses the decay rates to some extent ($|a_2|^2 \sim 7\%$). Note that $\Omega_{\alpha\beta\delta}^{XY} \propto (m_X - m_Y)$ comes from the application of a constituent $U(2,2)$ quark model [8], and this may lead to large suppression and uncertainty if $m_X$ and $m_Y$ are comparable in magnitude (e.g., $B_c^- \rightarrow K^- K^0$ and $\pi^-\pi^0$). In comparison with our rough results for $B_u^- \to D_s^- K^0$ and $\bar{B}_d^0 \to D_s^+ K^-$, the existing data give $\mathrm{Br}(B_u^- \to D_s^- K^0) < 1.1 \times 10^{-3}$ and $\mathrm{Br}(\bar{B}_d^0 \to D_s^+ K^-) < 2.4 \times 10^{-4}$ [1, 5].

The above discussions do not take into account the rescattering effect of final states due to strong interactions. Such effects may give rise to mixing of a "pure" decay mode (via a single quark diagram) with others, including those which were not originally coupled to this weak channel. Thus it is necessary to estimate the magnitude of final-state rescattering for those transitions listed in Eqs. (1) and (2). For illustration, here we take $B_u^- \to D^- \bar{K}^0$ for example to demonstrate that significant channel mixing can completely ruin a "pure" decay mode.

An isospin analysis shows that $B_u^- \to D^- \bar{K}^0$ may mix under rescattering with $B_u^- \to \bar{D}^0 K^-$ and $\bar{B}_d^0 \to \bar{D}^0 \bar{K}^0$ [13, 14] [3]. Note that $\bar{D}^0 \bar{K}^0$ is a pure $I = 1$ state, and $\bar{B}_d^0 \to \bar{D}^0 \bar{K}^0$ occurs only through a single color-mismatched spectator diagram [3, 14]. In contrast, $B_u^- \to \bar{D}^0 K^-$ takes place via both the color-mismatched spectator graph and the annihilation one. Ignoring final-state interactions, the amplitude of $\bar{B}_d^0 \to \bar{D}^0 \bar{K}^0$ can be factorized as:

$$\langle \bar{D}^0 \bar{K}^0 | \mathcal{H}_{\rm eff} | \bar{B}_d^0 \rangle \;=\; \frac{G_F}{\sqrt{2}}\, a_2 \, (V_{ub} V_{cs}^*) \, \Pi_{ucs}^{\bar{D}^0 \bar{K}^0} \;, \tag{7}$$

where the hadronic matrix element $\Pi_{ucs}^{\bar{D}^0 \bar{K}^0}$ is obtainable from the generic formula

$$\begin{aligned} \Pi^{XY}_{\alpha\beta\sigma} &\equiv \langle X | (\bar{\alpha}\beta)_{V-A} | 0 \rangle \langle Y | (\bar{\sigma} b)_{V-A} | B_\delta \rangle \\ &= -{\rm i} \left( m^2_{B_\delta} - m^2_Y \right) f_X \; F_0^{B_\delta Y} (m^2_X) \end{aligned} \tag{8}$$

for $B_\delta(b\bar{\delta}) \to X(\alpha\bar{\beta}) + Y(\sigma\bar{\delta})$. Under isospin invariance, one finds $\Pi_{ucs}^{\bar{D}^0 \bar{K}^0} = \Pi_{ucs}^{\bar{D}^0 K^-}$ and $\Omega_{scu}^{\bar{D}^0 K^-} = \Omega_{scu}^{D^- \bar{K}^0}$. After taking into account the rescattering effect of $D^- \bar{K}^0$, $\bar{D}^0 \bar{K}^0$ and $\bar{D}^0 K^-$, the transition amplitude of $B_u^- \to D^- \bar{K}^0$ can be written as

$$\langle D^- \bar{K}^0 | \mathcal{H}_{\rm eff} | B_u^- \rangle \;=\; \frac{G_F}{2\sqrt{2}} \, (V_{ub} V_{cs}^*) \left[ \left( a_2 \, \Pi_{ucs}^{\bar{D}^0 K^-} + 2 a_1 \, \Omega_{scu}^{D^- \bar{K}^0} \right) e^{{\rm i}\phi_0} \;-\; a_2 \, \Pi_{ucs}^{\bar{D}^0 K^-} e^{{\rm i}\phi_1} \right] \;, \tag{9}$$

where $\phi_0$ and $\phi_1$ are the strong phases of $I = 0$ and $I = 1$ states. It is obvious that contribution of the rescattering term $\Pi_{ucs}^{\bar{D}^0 K^-}$ to $\langle D^- \bar{K}^0 | \mathcal{H}_{\rm eff} | B_u^- \rangle$ disappears if $\Delta\phi \equiv \phi_1 - \phi_0 = 0$ [14]. As a result, we find the effect of nonvanishing $\Delta\phi$ on the branching ratio of $B_u^- \to D^- \bar{K}^0$:

$$\begin{aligned} {\rm R}(\Delta\phi) &\equiv \frac{{\rm Br}(B_u^- \to D^- \bar{K}^0)|_{\Delta\phi \neq 0}}{{\rm Br}(B_u^- \to D^- \bar{K}^0)|_{\Delta\phi = 0}} \\ &= 1 \;-\; \xi \; \sin(\Delta\phi) \;+\; \xi^2 \sin^2 \left( \frac{\Delta\phi}{2} \right) \end{aligned} \tag{10}$$

[3] Here a reasonable assumption is that no additional channel mixes with these three modes.

with

$$\xi \;\equiv\; \pm \, \frac{a_2}{a_1} \left| \frac{\Pi^{\bar{D}^0 K^-}_{ucs}}{\Omega^{D^- \bar{K}^0}_{scu}} \right| \; . \tag{11}$$

The sign ambiguity of $\xi$ arises from the unknown relative sign between $\Pi^{\bar{D}^0 K^-}_{ucs}$ and $\Omega^{D^- \bar{K}^0}_{scu}$. The size of $\Pi^{\bar{D}^0 K^-}_{ucs}$ can be estimated with the inputs $F^{B_d K}_0(0) = 0.38$ and $f_{D^0} = 0.253$ GeV [6, 12]. We approximately obtain $\xi \approx \pm 18.8$. The change of $\mathrm{R}(\Delta\phi)$ as a function of $\Delta\phi$ is numerically illustrated in Fig. 2.

It is clear that the significant rescattering effect ($|\Delta\phi| \geq 50^0$) can dramatically enhance the branching ratio of $B^-_u \rightarrow D^- \bar{K}^0$ to the level $O(10^{-6})$. In this case, the transition is indeed dominated by the contribution from $\bar{B}^0_d \rightarrow \bar{D}^0 \bar{K}^0$ and $B^-_u \rightarrow \bar{D}^0 K^-$. Considering $DK$ scattering via a $t$-channel exchange of Regge trajectories, Deshpande and Dib have estimated the strong phase shift $\Delta\phi$ and obtained $\tan(\Delta\phi) \approx -0.14$ [14]. This result has a two-fold ambiguity: for $\Delta\phi \approx -8^0$, $\mathrm{R}(\Delta\phi)$ deviates only a little from $\mathrm{R}(0)$; for $\Delta\phi \approx 172^0$, $\mathrm{R}(\Delta\phi)/\mathrm{R}(0) \sim 10^2$ may turn out. Note also that $\mathrm{R}(\Delta\phi)$ is insensitive to the sign of $\xi$, due to the fact $|\xi| >> 1$.

Certainly the above calculation approaches have many uncertainties which are unable to be removed to the limit of our present understanding of the $W$-exchange and annihilation transitions. Thus the relevant quantitative results might not be trustworthy, but only serve as illustration of the possible qualitative effects. A reliable examination of the true role of $W$-exchange and annihilation quark diagrams playing in different types of hadronic rare $B$ decays deserves further theoretical and experimental efforts.

I am grateful to H. Fritzsch for his warm hospitality. I am also indebted to C.O. Dib for his patient help in understanding Ref. [14] and to Y.Q. Chen for a useful conversation. This research was supported by the Alexander von Humboldt Foundation of Germany.

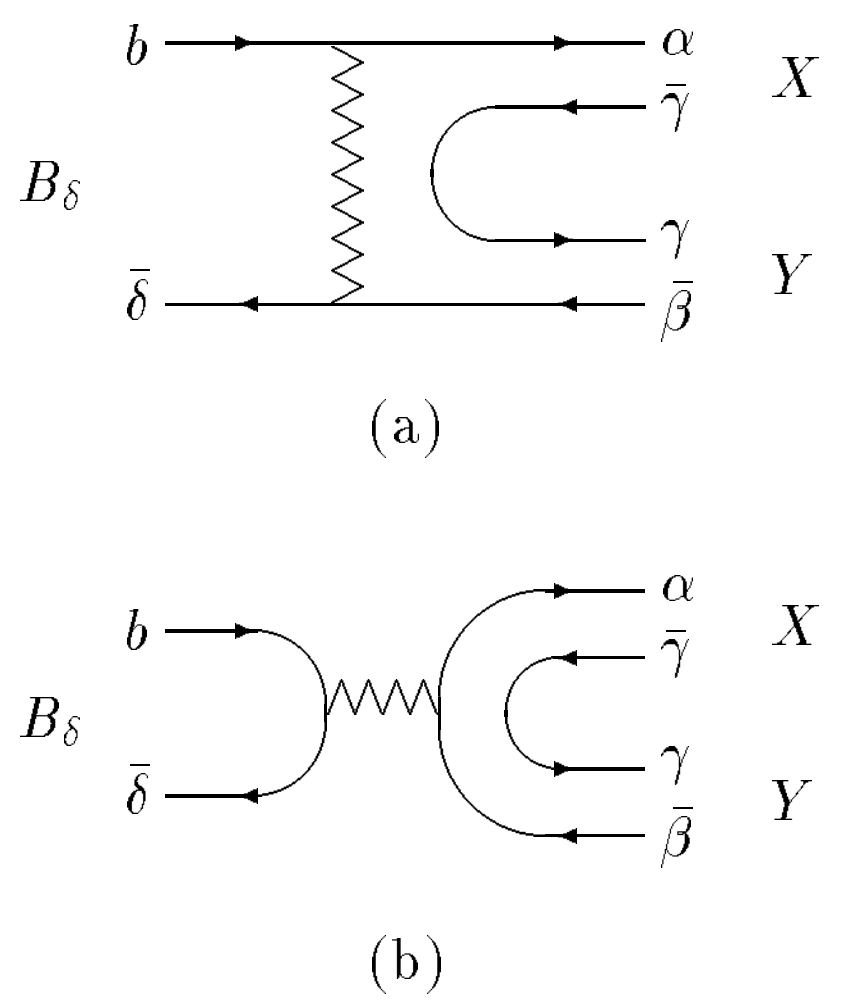

Figure 1: A graphic description of the two-body mesonic decay $B_\delta(b\bar{\delta}) \rightarrow X(\alpha\bar{\gamma}) + Y(\gamma\bar{\beta})$: (a) the $W$-exchange diagram with $\delta = d$ or $s$, $\alpha = u$ or $c$, $\beta = c$ or $u$, and $\gamma = u$, $d$ or $s$; (b) the annihilation diagram with $\delta = u$ or $c$, $\alpha = d$ or $s$, $\beta = u$ or $c$, and $\gamma = u$, $d$ or $s$.

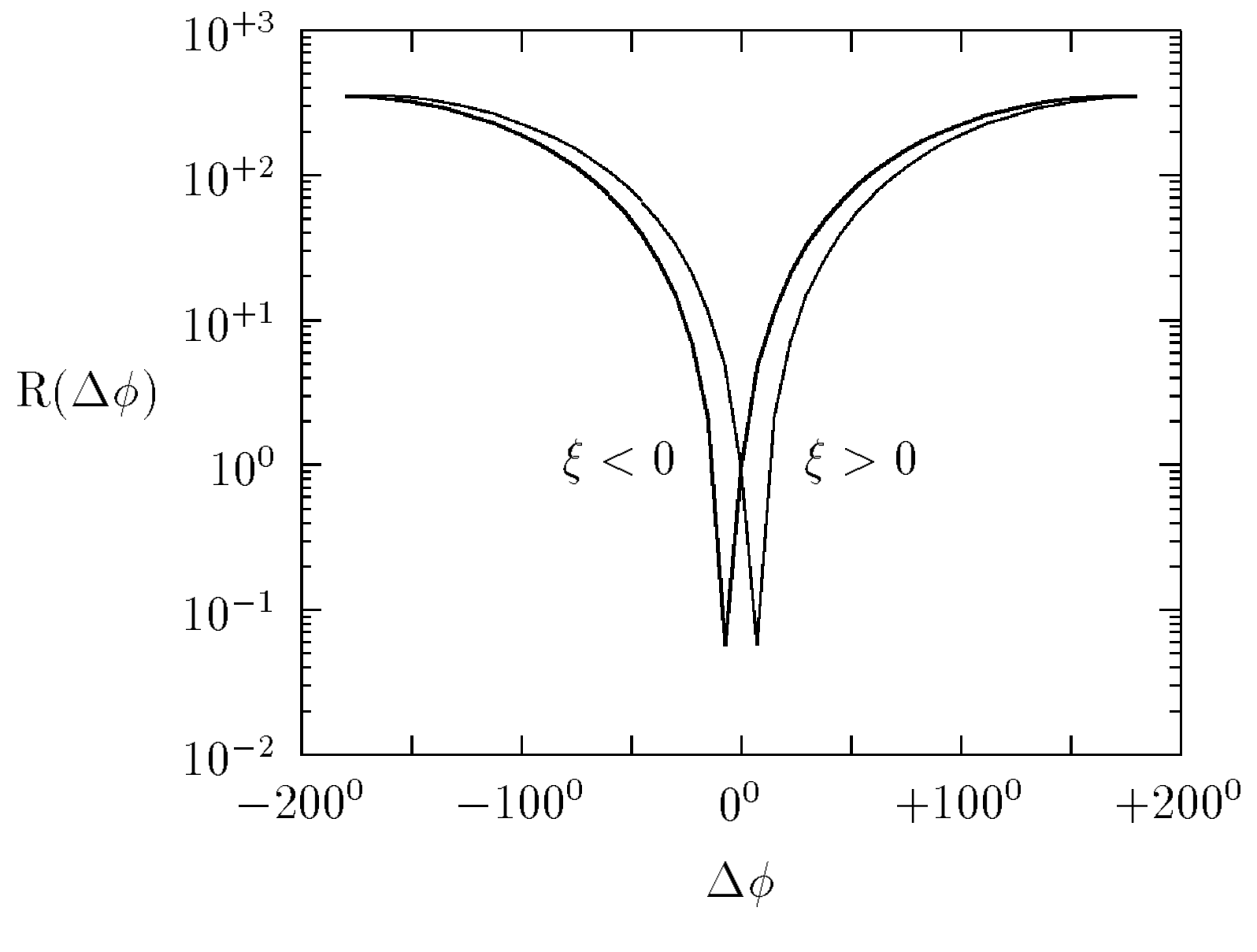

Figure 2: The change of R($\Delta\phi$) as a function of the rescattering phase shift $\Delta\phi$.